\documentclass[aps,prl,twocolumn,letterpaper,superscriptaddress,showpacs]{revtex4-2}
\usepackage{keyval}%
\usepackage{graphicx}
\usepackage{dcolumn}
\usepackage{bm}
\usepackage{color}
\usepackage{amsmath,amssymb}

\begin{document}

\title{The hidden competing phase revealed by first-principles calculations of phonon instability in the nearly optimally doped cuprate La$_{1.875}$Sr$_{0.125}$CuO$_4$}
\author{Chi-Cheng Lee}
\affiliation{Department of Physics, Tamkang University, Tamsui, New Taipei 251301, Taiwan}%
\affiliation{Research Center for X-ray Science, College of Science, Tamkang University, Tamsui, New Taipei 251301, Taiwan}%
\author{Ji-Yao Chiu}
\affiliation{Department of Physics, Tamkang University, Tamsui, New Taipei 251301, Taiwan}%
\author{Yukiko Yamada-Takamura}
\affiliation{School of Materials Science, Japan Advanced Institute of Science and Technology (JAIST), 1-1 Asahidai, Nomi, Ishikawa 923-1292, Japan}%
\author{Taisuke Ozaki}
\affiliation{Institute for Solid State Physics, The University of Tokyo, 5-1-5 Kashiwanoha, Kashiwa, Chiba 277-8581, Japan}%
\date{\today}

\begin{abstract}
The representative cuprate, La$_{2-x}$M$_x$CuO$_4$, with M = Sr and $x = 1/8$ is studied via first-principles calculations in the 
high-temperature tetragonal (HTT), low-temperature orthorhombic (LTO), and low-temperature less-orthorhombic (LTLO) structures.
By suppressing the magnetism and superconductivity, the LTLO phase, which has rarely been observed in La$_{2-x}$Sr$_x$CuO$_4$, 
is found to be the ground state, where the structural phase transitions, HTT$\rightarrow$LTO$\rightarrow$LTLO, 
can be understood via phonon instability. While the La-O composition is identified to be responsible for the phonon softening, 
the superconducting CuO$_2$ layer is dynamically stable. The LTLO phase, which can exhibit a $\sim$20 meV splitting 
in the density of states, is proposed to have an intimate relationship with the observed pseudogap and the charge density wave giving the stripe. 
We argue that at low temperatures, the superconducting LTO La$_{1.875}$Sr$_{0.125}$CuO$_4$ competes with the phonon-preferred LTLO phase 
by spontaneously forming the Cooper pairs, resulting in suppressing the stripe. Therefore, the revealed LTLO phase is indispensable 
for understanding La$_{2-x}$Sr$_x$CuO$_4$.
\end{abstract}
  
\maketitle

The observation of the superconductivity at 39 K in MgB$_2$\cite{Nagamatsu} set a milestone for the superconducting transition temperature (T$_c$)
in phonon-mediated conventional superconductors.\cite{PhysRev.108.1175} Recently, the highest T$_c$ in the conventional superconductors was
renewed again reaching 203 K in H$_3$S at high pressure.\cite{Drozodov,Duan} Very recently, the room-temperature superconductor, namely, superconducting at
288 K, has been experimentally realized in carbonaceous sulfur hydride at 267 GPa,\cite{Snider} which remarks a great progress in the long search for the
room-temperature superconductors under ambient conditions. Surprisingly, the underlying mechanism for the hydrogen-rich materials is still governed by the 
electron-phonon coupling,\cite{Kim2793,PhysRevB.96.100502} proposed by Bardeen, Cooper, and Schrieffer (BCS) six decades ago.\cite{PhysRev.108.1175} 
On the other hand, the cuprate family, the representative of unconventional superconductors, is known for containing rich strong-correlation physics, 
where the theoretical prediction of T$_c$ based on the electron-phonon pairing mechanism alone has not been successful in explaining the observed 
high T$_c$.\cite{RevModPhys.75.473,RevModPhys.78.17,plakida2010high} Given that the Hg-based cuprates have been keeping the record of the highest T$_c$ 
at ambient conditions,\cite{Schilling} the cuprate family is still on the list of the promising candidates for realizing room-temperature superconductivity 
without high pressure. 

So far the key for realizing the room-temperature T$_c$ in cuprates has not been found, but not all the properties of superconductivity 
in cuprates are too elusive to understand via the conventional mechanisms.\cite{PhysRevLett.119.237001} Given that the electron-phonon coupling is important 
for the carbonaceous sulfur hydride\cite{Snider} and first-principles calculations have suggested the possession of strong electron-phonon coupling in 
cuprates,\cite{PhysRevB.47.1002,PhysRevX.3.021011} it is interesting to ask whether or not something related to the phonon properties has been missing in the studies of cuprates. 
In cuprates, high-temperature tetragonal (HTT), low-temperature tetragonal (LTT), 
low-temperature orthorhombic (LTO), and low-temperature less-orthorhombic (LTLO) structures are ubiquitously observed, 
for example, in La$_{2-x}$Ba$_x$CuO$_4$.\cite{PhysRevB.83.104506}
Generally speaking, the structural phase transitions can be explained by a soft-phonon model,\cite{PhysRevLett.62.2751,cherny1995soft,Kimura} 
such as the first-order LTO$\rightarrow$LTT and the second-order HTT$\rightarrow$LTO$\rightarrow$LTLO$\rightarrow$LTT transitions.
But the actual transitions depend on the studied cuprates. 

First-principles calculations without using advanced functionals cannot reproduce the observed insulating gaps in the undoped 
cuprates\cite{furness2018accurate} but are capable of describing the phonon properties in the metallic phases.\cite{PhysRevLett.62.831,PhysRevB.47.1002,PhysRevB.59.9278,TAVANA201520}
For the representative cuprate, La$_{2-x}$M$_x$CuO$_4$, La$_{2-x}$Ba$_x$CuO$_4$ exhibits rich phases and an anomalously deep depression at the superconducting phase boundary 
around $x=1/8$.\cite{PhysRevB.83.104506,PhysRevLett.88.167008} In contrast, La$_{2-x}$Sr$_x$CuO$_4$ exhibits a more standard dome-shaped boundary, and 
the HTT and LTO phases are the only major players in the phase diagram.\cite{Day_1987,PhysRevB.40.2254,PhysRevB.46.14034,PhysRevB.50.3221,PhysRevB.50.13125,PhysRevB.89.224513} 
The HTT$\rightarrow$LTO transition can be understood via the phonon softening in the HTT phase.\cite{PhysRevB.38.185,PhysRevB.59.9278,PhysRevLett.62.2751,TAVANA201520} 
La$_{2-x}$Sr$_x$CuO$_4$ is then served as a relatively simple system for investigation. Nevertheless,
the missing LTLO phase was observed at a seemingly exclusive doping level, namely, $x=0.12$.\cite{PhysRevB.51.9045,PhysRevB.61.11922}
Recently, the presence of the LTLO structure, which is intimately related to the charge stripe order, has also been evidenced in La$_{1.93}$Sr$_{0.07}$CuO$_4$ 
by neutron scattering experiments.\cite{PhysRevB.92.174525}
In this study, we will address the phonon instability in the nearly optimally doped LTO La$_{1.875}$Sr$_{0.125}$CuO$_4$ based on the first-principles supercell calculations, 
which are different from previously adopted rigid-band and virtual-crystal approximations,\cite{PhysRevB.47.1002,TAVANA201520}
and demonstrate that the LTLO phase is the ground state by quenching the magnetism, 
large-scale charge ordering,\cite{PhysRevLett.76.3412,PhysRevB.89.224513,wang2019real} and superconductivity.

\begin{figure}[tbp]
\includegraphics[width=1.00\columnwidth,clip=true,angle=0]{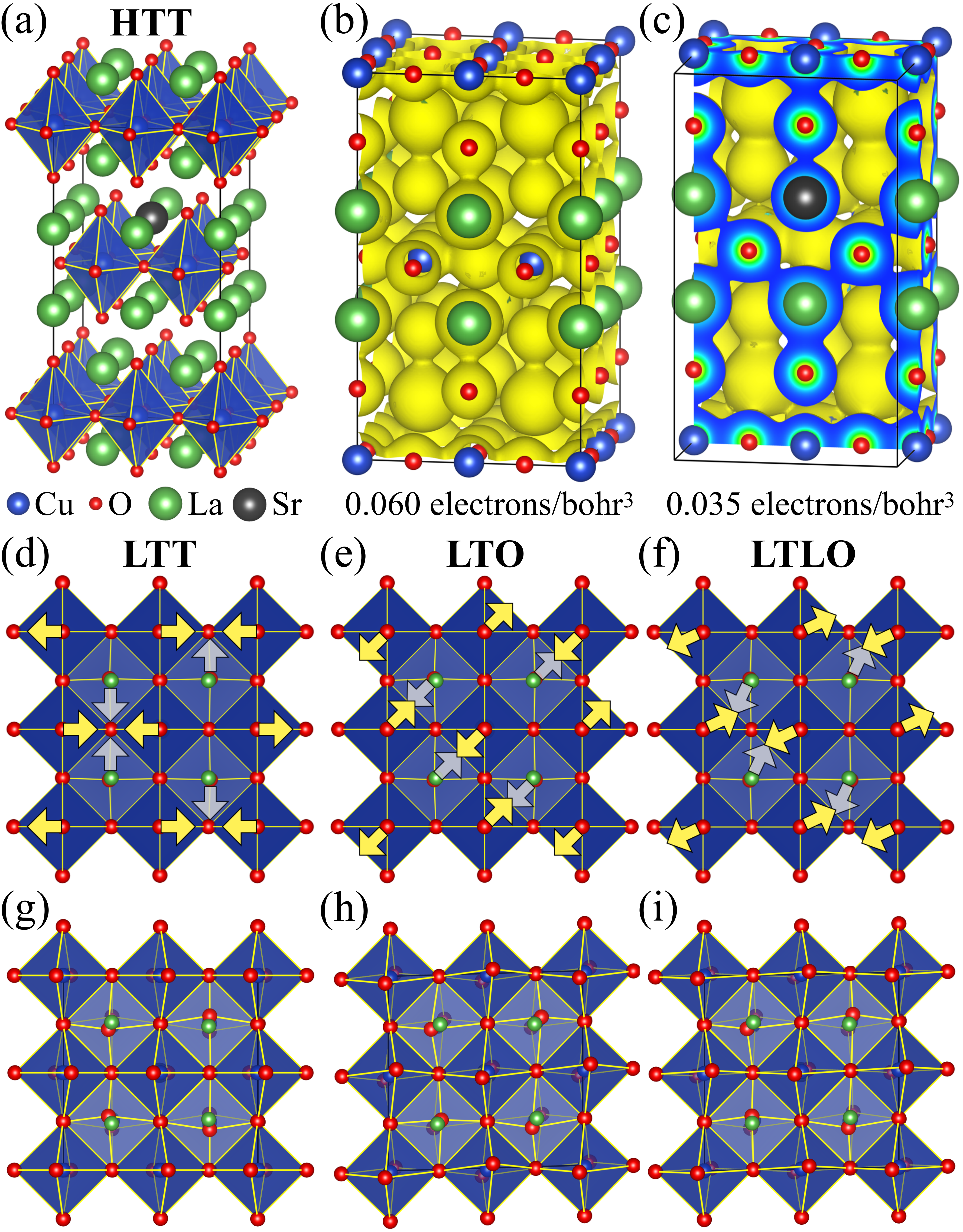}
\caption{(a) Side view of the high-temperature tetragonal (HTT) structure of La$_{1.875}$Sr$_{0.125}$CuO$_4$, where $a=7.62$\AA~and $c=13.22$\AA. 
The CuO$_6$ is presented by the octahedron. 
The isosurfaces of the charge density at 0.060 and 0.035 electrons/bohr$^3$ are presented in (b) and (c), respectively.
The tilting directions of the apical O atoms are indicated by the yellow (top layer) and grey (middle layer) arrows from the top view to describe the derived 
(d) low-temperature tetragonal (LTT), (e) low-temperature orthorhombic (LTO), and (f) low-temperature less orthorhombic (LTLO)
structures from the HTT one. The calculated tilted octahedra of LTT, LTO, and LTLO structures are 
shown in (g), (h), and (i), respectively. The plots were generated using VESTA.\cite{VESTA}
}
\label{fig:struct}
\end{figure}

The first-principles calculations based on the density functional theory (DFT) were performed using the OpenMX code,\cite{openmx} where the 
generalized gradient approximation (GGA), norm-conserving pseudopotentials, and optimized pseudo-atomic basis functions were adopted.\cite{GGA,Morrison,Ozaki} 
Three, three, three, and two optimized radial functions were allocated for the $s$, $p$, $d$, and $f$ orbitals, respectively, for each 
La atom with a cutoff radius of 8 bohr, denoted as La8.0-s3p3d3f2. For the Sr, Cu, and O atoms, Sr10.0-s3p2d2f2, Cu6.0-s2p2d2, and 
O7.0-s2p2d1 were adopted, respectively. A cutoff energy of 500 Ry was used for numerical integrations and for the solution of 
the Poisson equation. La$_{1.875}$Sr$_{0.125}$CuO$_4$ was studied via the unit cell containing a Sr atom, 8 Cu atoms, 15 La atoms, and 32 O atoms,
where one of the symmetrically equivalent La atoms was replaced by the Sr atom from the 
HTT, LTT, LTO, and LTLO structures with $I4/mmm$, $P4_2/ncm$, $Bmab$, and $Pccn$ symmetry, respectively. 
After the substitution, the structures were fully relaxed ($P1$) within DFT-GGA using a $4\times4\times2$ $k$-point sampling 
and all the forces are less than $10^{-4}$ Ha/bohr. The relaxed tilted octahedra are presented in Fig.~\ref{fig:struct}.\cite{supp}
The force constants needed for constructing the dynamical matrix were obtained from the ($2\times2\times1$) supercell calculations (224 atoms). 

\begin{figure}[tbp]
\includegraphics[width=1.00\columnwidth,clip=true,angle=0]{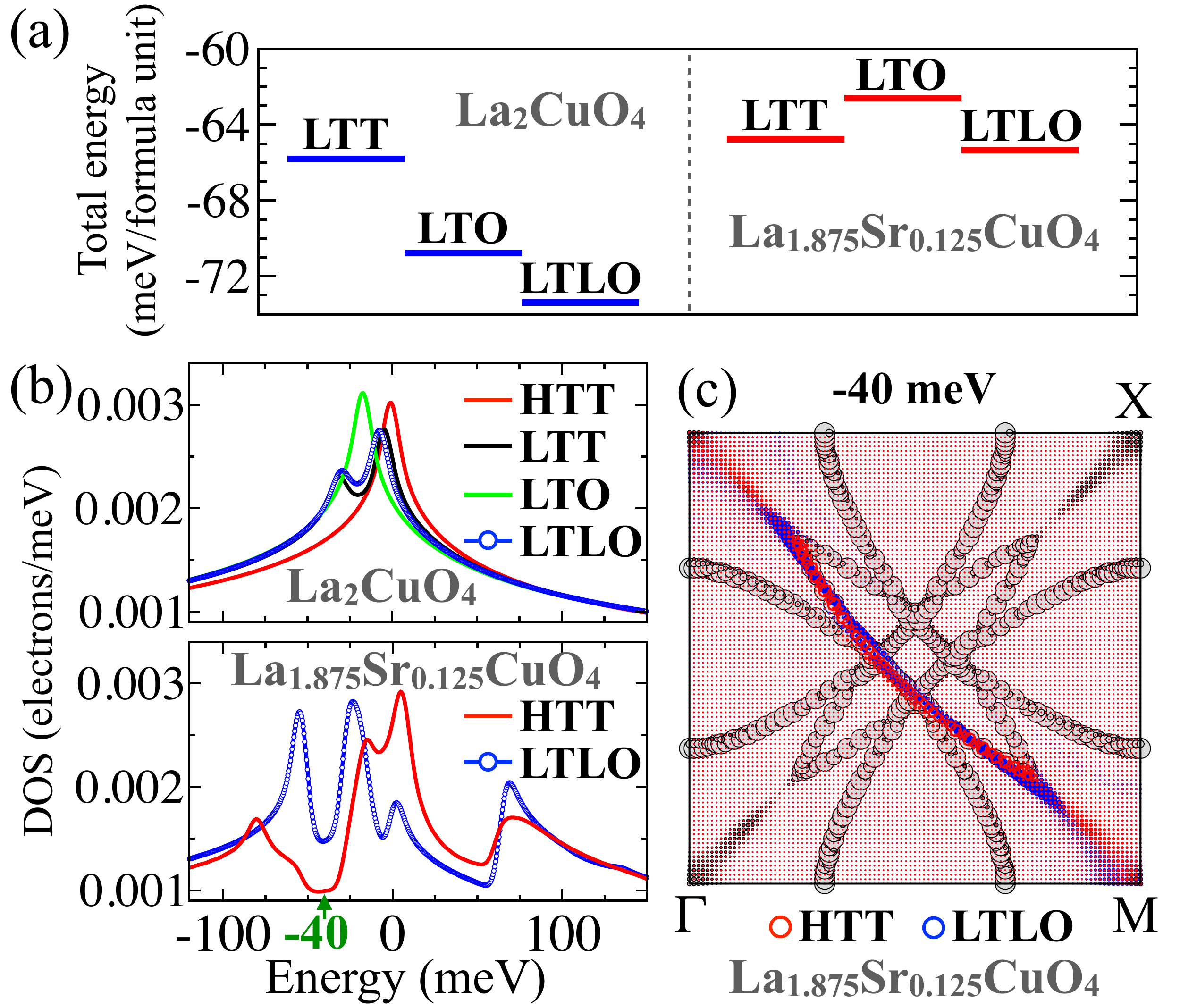}
\caption{(a) DFT-GGA energy levels of LTT, LTO, and LTLO phases relative to the total energy of HTT phase for La$_{2}$CuO$_4$
and La$_{1.875}$Sr$_{0.125}$CuO$_4$, respectively. (b) Density of states (DOS) per formula unit for $k_z=0$ near the Fermi level (0 meV) 
in La$_{2}$CuO$_4$ and La$_{1.875}$Sr$_{0.125}$CuO$_4$.
(c) Spectral weight of Kohn-Sham orbitals in HTT (red circles) and LTLO (blue circles) La$_{1.875}$Sr$_{0.125}$CuO$_4$ unfolded to 
the Brillouin zone of the primitive cell (7 atoms) with M:(0.5,0,0) and X:(0.5,0.5,0) in units of the reciprocal lattice vectors of 
the conventional cell (14 atoms). The original (folded) weight in HTT phase is presented by black circles.
}
\label{fig:energy}
\end{figure}

The untilted CuO$_6$ octahedra in the HTT structure, shown in Fig.~\ref{fig:struct} (a), serve as the building blocks to 
construct the LTT, LTO, and LTLO structures with different tilts, which are indicated by the arrows in Figs.~\ref{fig:struct} (d), (e), and (f), respectively. 
The calculated structures within DFT-GGA are shown in Figs.~\ref{fig:struct} (g), (h), and (i), respectively, where the expected tilts can be recognized
even with the presence of the Sr atom. The LTLO structure is apparently a mixture of the LTO and LTT ones, which supports a continuous LTLO-LTO or LTLO-LTT 
phase transition, whereas the direct LTO-LTT phase transition is first-order.\cite{plakida2010high} Although the HTT and LTO phases are the only major players 
in La$_{1-x}$Sr$_{x}$CuO$_4$, the total energies of the LTT, LTO, and LTLO phases relative to the HTT phase within DFT-GGA presented 
in Fig.~\ref{fig:energy} (a) reveal that the LTLO phase has the lowest energy. At $x=0$, DFT-GGA without considering magnetism cannot well describe LaCuO$_4$, 
but for $x=0.125$, the long-range antiferromagnetic order has been destroyed so that DFT-GGA provides a good description for the normal-state 
La$_{1.875}$Sr$_{0.125}$CuO$_4$. In Fig.~\ref{fig:energy} (b), the density of states is presented and 
exhibits a split feature in the LTLO and LTT La$_2$CuO$_4$. 
The split van Hove singularities near the Fermi level can offer an explanation for the presence of the pseudogaps based on the dynamic Jahn-Teller effect.\cite{markiewicz}   
The splitting can be further introduced and enhanced in La$_{1.875}$Sr$_{0.125}$CuO$_4$ due to the affected CuO$_6$ tilting and Brillouin-zone periodicity. 
However, the intensity of the two peaks forming the gap at -40 meV is more prominent in the LTLO phase than that in the HTT phase. 
The Fermi arc\cite{PhysRevB.74.224510} can be realized by unfolding the spectral weight\cite{Lee_2013} to a larger Brillouin zone at this energy,
as illustrated in Fig.~\ref{fig:energy} (c).

\begin{figure}[tbp]
\includegraphics[width=1.00\columnwidth,clip=true,angle=0]{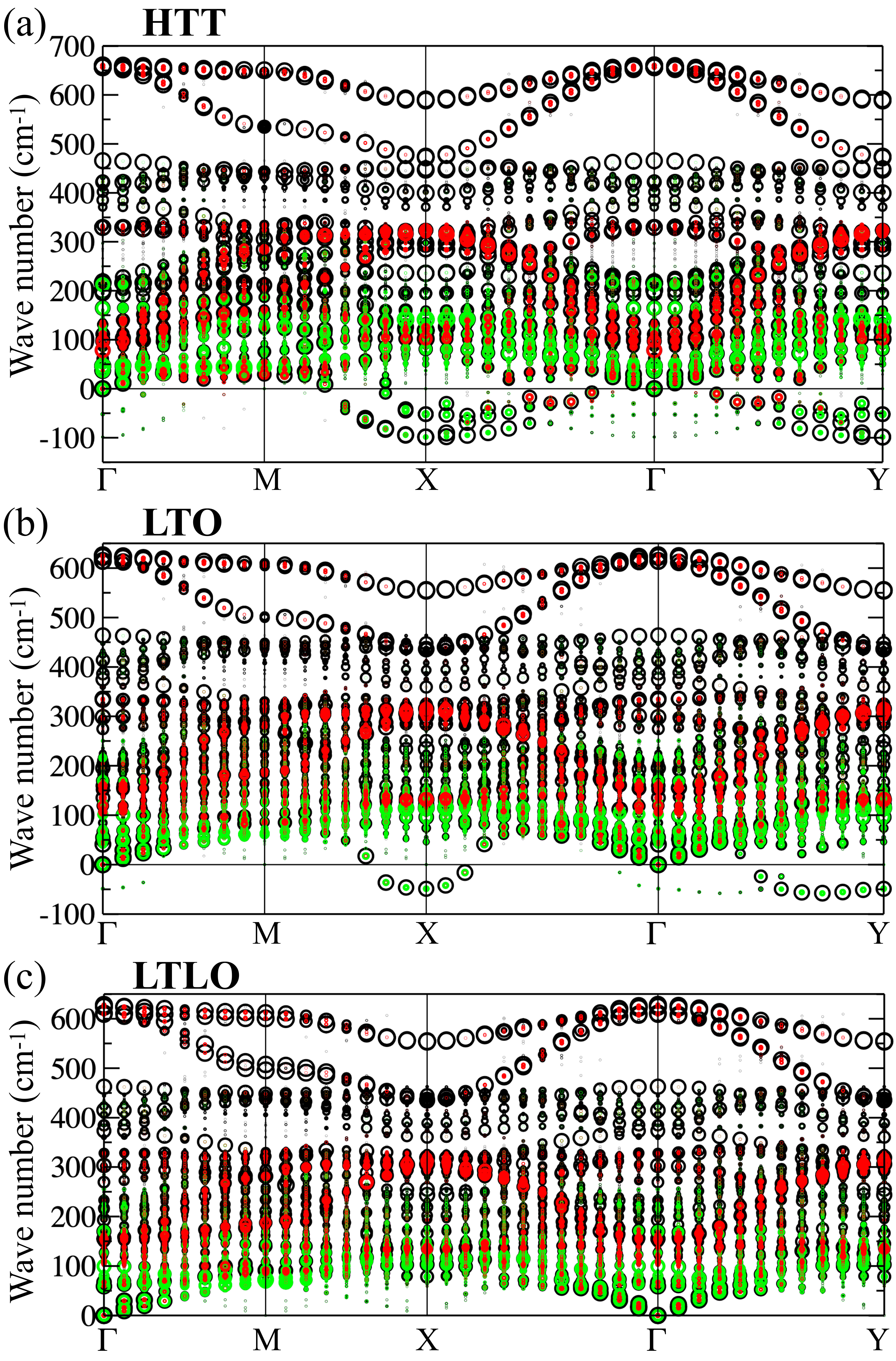}
\caption{DFT-GGA phonon dispersions in (a) HTT, (b) LTO, and (c) LTLO La$_{1.875}$Sr$_{0.125}$CuO$_4$. The black circle represents the unfolded weight
for each vibrational mode in the Brillouin zone of the conventional cell (14 atoms) with M:(0.5,0,0), X:(0.5,0.5,0), and Y:(-0.5,0.5,0).
The radii of red (green) circles are proportional to the Cu (La) contribution.
}
\label{fig:dispersion}
\end{figure}

To verify the dynamical stability of the HTT, LTO, and LTLO phases, the phonon properties are investigated and discussed here. 
The phonon dispersion in the HTT La$_{1.875}$Sr$_{0.125}$CuO$_4$ is presented in Fig.~\ref{fig:dispersion} (a). One of the two degenerate vibrational modes 
at the lowest (imaginary) frequency at X is in accordance with the guiding arrows for the HTT$\rightarrow$LTO transition, shown in Fig.~\ref{fig:struct} (e), 
and the other mode has a 90-degree rotation. Presumably, the instability can be removed by following either one of the mutually orthogonal sets of eigendisplacements. 
To further verify the stability of the LTO phase, the phonon dispersion is presented in Fig.~\ref{fig:dispersion} (b). It can be found that
imaginary-frequency modes still exist. Interestingly, the lowest-frequency vibrational mode at X is in accordance with 
the previously mentioned 90-degree-rotation mode. This indicates that the system can be further stabilized via additional tilting guided by 
the persisting phonon instability, 
which brings the LTO structure to the LTLO one. Finally, the phonon dispersion in the LTLO phase is shown in Fig.~\ref{fig:dispersion} (c), where 
no imaginary frequencies are present. The soft-phonon picture advocates the HTT$\rightarrow$LTO$\rightarrow$LTLO transitions in the normal-state 
La$_{1.875}$Sr$_{0.125}$CuO$_4$ and is consistent with the calculated total energies. 

The observed pseudogap in La$_{2-x}$Sr$_x$CuO$_4$ can be related to the LTLO phase by considering the atoms displacing in a double well connecting
the LTO and LTLO structures. As already illustrated in Fig.~\ref{fig:energy} (b), the split density of states in the LTLO structure could be responsible for the pseudogap.
At lower temperatures, the structure intends to deviate from the center of the double well and could develop charge-density-wave fluctuations.\cite{PhysRevB.89.224513,wang2019real}
Since the LTLO phase may experience a continuous transition to the LTO or LTT phase, the observed stripe showing a mixture of LTO-like 
and LTT-like distortions\cite{PhysRevLett.76.3412} could have an intimate relationship with the LTLO structure.
Our result also supports the mentioned role of the LTLO structure for the observed spin and charge fluctuations at high temperatures.\cite{PhysRevB.92.174525}
The LTT La$_{1.875}$Sr$_{0.125}$CuO$_4$, which is also dynamically stable, has only slightly higher total energy than the LTLO one. 
The almost degenerate energy could also play a role for the depression around $x=0.125$ at the superconducting phase boundary.\cite{PhysRevB.89.224513}
Regarding that the energy difference between the LTO and LTLO phases is comparable with the superconducting gap, it is reasonable to argue that the absence of the LTLO phase 
at lower temperatures is due to the electron pairing that stabilizes the superconducting LTO phase in competing with the phonon-preferred LTLO phase.

We now reveal the major player for the phonon instability in the HTT and LTO phases. 
As presented by the radii of circles in Fig.~\ref{fig:dispersion} (a) for the HTT phase, both the Cu and La atoms are found 
to contribute to the imaginary-frequency branches. More specifically, the Cu atoms mainly contribute to the shallower part, 
whereas the La atoms are responsible for almost all the imaginary-frequency modes. For the most unstable degenerate optical modes, whose frequency
is -98.5 $cm^{-1}$ at X, the atomic contribution is mainly from the O atoms. In the LTO phase, the Cu contribution  
disappears significantly in the imaginary-frequency branches, as shown in Fig.~\ref{fig:dispersion} (b),
suggesting that the ingredients of phonon instability are composed of the La and O atoms. 
This finding implies that the structural phase transitions are driven by the La-related bonding. In fact, the isosurfaces of charge density in
HTT La$_{1.875}$Sr$_{0.125}$CuO$_4$, shown in Figs.~\ref{fig:struct} (b) and (c) at 0.060 and 0.035 electrons/bohr$^3$, respectively, clearly show that 
the apical O atoms only weakly bond to the center Cu atoms but strongly bond to the La atoms. Therefore, the intuitively understandable tilting of CuO$_6$ 
octahedra, as shown in Fig.~\ref{fig:struct}, does not describe the underlying mechanism for the structural phase transitions. 

\begin{figure}[tbp]
\includegraphics[width=1.00\columnwidth,clip=true,angle=0]{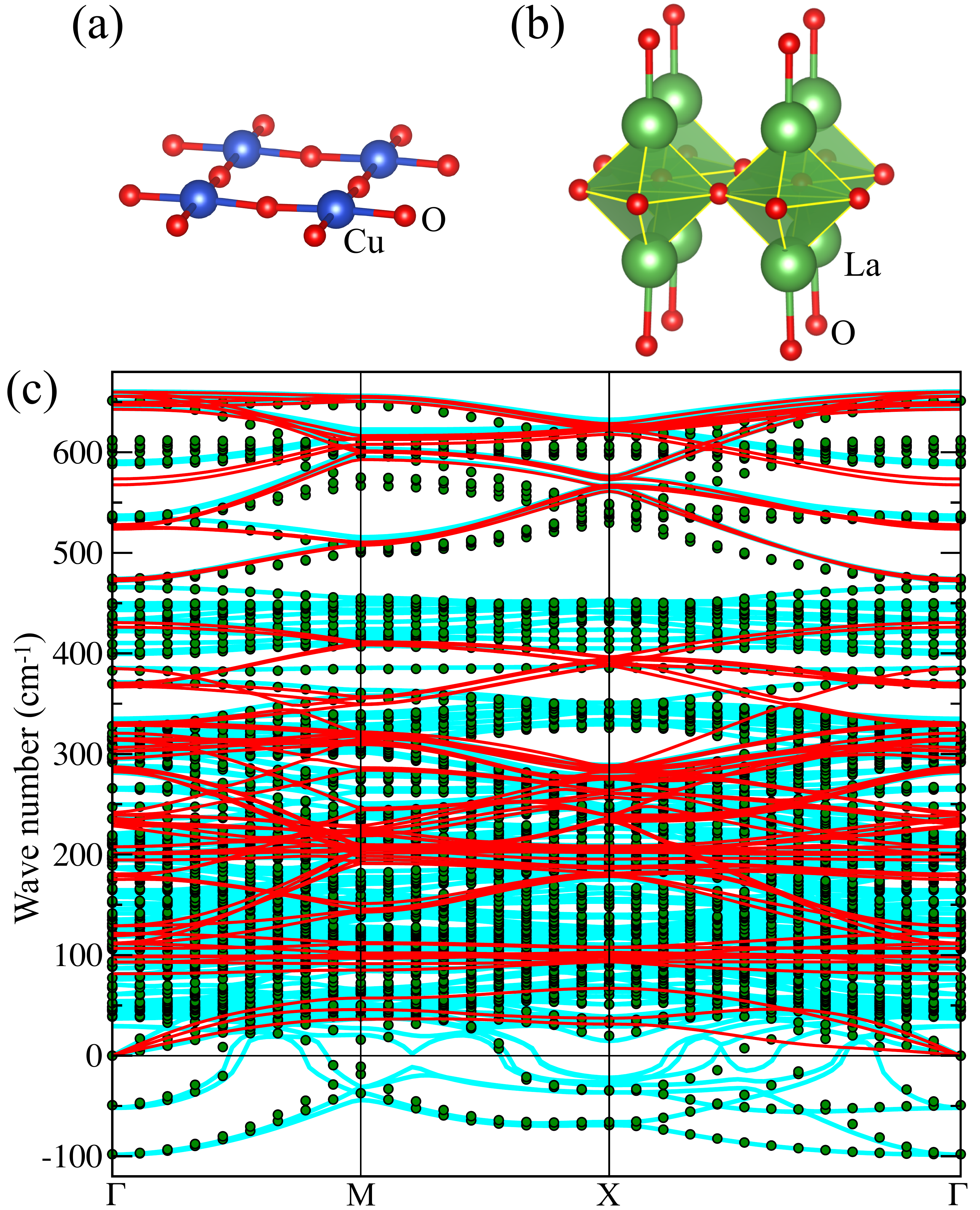}
\caption{(a) The CuO$_2$ layer and (b) the extended octahedra composed of La and O atoms in HTT La$_{1.875}$Sr$_{0.125}$CuO$_4$.
(c) Reconstructed phonon dispersions of the (a) planar CuO$_2$ and (b) extracted octahedra are plotted by red curves and green circles, respectively. 
The original supercell phonon dispersion with M:(0.5,0,0) and X:(0.5,0.5,0) is presented by cyan curves.
}
\label{fig:CuvsLa}
\end{figure}

To demonstrate the roles of partial compositions in HTT La$_{1.875}$Sr$_{0.125}$CuO$_4$ for the phonon instability, 
such as the structural stability of the CuO$_2$ plane shown in Fig.~\ref{fig:CuvsLa} (a), we can rediagonalize    
the dynamical matrix by treating the irrelevant atoms in the unit cell as a rigid body without internal degrees of freedom of vibration, 
which is in accordance with the mobile block Hessian approximation.\cite{Ghysels,Terrett} 
The force constants between the rigid clusters in the different unit cells are set to zero. The resultant dispersion of 
the CuO$_2$ plane is shown in Fig.~\ref{fig:CuvsLa} (c), where no imaginary frequencies can be found. 
Alternatively, we can consider all the Cu and Sr atoms in the unit cell forming a heavy rigid cluster. 
As presented in Fig.~\ref{fig:CuvsLa} (c), the lowest-frequency branches in HTT La$_{1.875}$Sr$_{0.125}$CuO$_4$ are well reproduced. 
This provides a solid support that the extracted structure composed of the La and O atoms, as shown in Fig.~\ref{fig:CuvsLa} (b), can well describe 
the phonon instability in the HTT phase. 

In conclusion, the phonon properties in the nearly optimally doped La$_{1.875}$Sr$_{0.125}$CuO$_4$ have been studied within DFT-GGA, where 
the La-O composition is identified to be responsible for the phonon instability giving rise to the HTT$\rightarrow$LTO$\rightarrow$LTLO transitions,
while the superconducting CuO$_2$ layer itself is dynamically stable.
One of the softest doubly degenerate modes in the HTT phase triggers the HTT$\rightarrow$LTO transition,
and the other mode still persists in the LTO phase to further drive the LTO$\rightarrow$LTLO transition.
The calculated total energies also support that the LTLO phase is the ground state at $x=0$ and $1/8$ by quenching the magnetism,
charge-density-wave order, and superconductivity.
We propose that the LTO structure can be described by a displacive model involving the LTLO$\leftrightarrow$LTO$\leftrightarrow$LTLO displacements. 
The exhibited $\sim$20 meV splitting in the density of states in the LTLO phase can be related to the observed pseudogap.
At low temperatures, we expect that the LTLO phase encounters a fierce competition with the observed LTO phase, which can gain more energy from the electron pairing 
in the superconducting state and therefore stabilize the LTO structure. This indicates that the calculation of T$_c$ via the electron-phonon coupling 
cannot be decoupled from the superconducting state. Finally, we emphasize that the missing LTLO phase provides an indispensable ingredient 
in describing La$_{2-x}$Sr$_x$CuO$_4$.

\begin{acknowledgments}
The calculations were carried out using the facilities in JAIST and Tamkang University.
Chi-Cheng Lee acknowledges the Ministry of Science and Technology of Taiwan for financial support under contract No. MOST 108-2112-M-032-010-MY2.
\end{acknowledgments}

\bibliography{refs}

\end{document}